\documentclass[aps,prb,twocolumn,superscriptaddress,showpacs]{revtex4-1}
\usepackage{amssymb,latexsym}
\usepackage{amsmath}
\usepackage{tensor}
\usepackage[utf8]{inputenc}
\usepackage[colorlinks=true,linkcolor=blue,citecolor=blue,urlcolor=blue]{hyperref}
\usepackage{graphicx}     
\usepackage[dvipsnames]{xcolor}
\usepackage{floatflt}     
\usepackage{wrapfig}
\usepackage{color}
\usepackage{pstricks}
\usepackage{subcaption}
\usepackage{color}
\definecolor{DarkRed}{rgb}{0.35,0.01,0.01}
 \definecolor{Linen}{rgb}{0.98,0.98,0.94}
 \definecolor{Blue}{rgb}{0.,0.,1.0}
 \definecolor{DarkBlue}{rgb}{0.099,0.099,0.44}
 \definecolor{DarkGreen}{rgb}{0.0,0.4,0.0}
 \definecolor{Turquoise}{rgb}{0.0,0.9,0.7}
 
\begin{document}

 \title{Exploration of The Duality Between Generalized Geometry and Extraordinary Magnetoresistance}
\author{Leo Rodriguez}
\email[]{rodriguezl@grinnell.edu}
\affiliation{Department of Physics, Grinnell College, Grinnell, IA 50112}
\author{Shanshan Rodriguez}
\email[]{rodriguezs@grinnell.edu}
\affiliation{Department of Physics, Grinnell College, Grinnell, IA 50112}
\author{Sathwik Bharadwaj}
\email[]{sathwik@wpi.edu}
\affiliation{Department of Physics, Worcester Polytechnic Institute, Worcester, MA 01609}
\author{L. R. Ram-Mohan}
\email[]{lrram@wpi.edu}
\affiliation{Department of Physics, Worcester Polytechnic Institute, Worcester, MA 01609}  
\date{\today}
\begin{abstract}
  We outline  the duality between the  extraordinary magnetoresistance
  (EMR), observed in semiconductor-metal hybrids, and non-symmetric
  gravity coupled to a diffusive $U(1)$ gauge field. The
  corresponding gravity theory  may be interpreted  as the generalized
  complex  geometry of  the  semi-direct  product of  the symmetric
  metric   and   the  antisymmetric Kalb-Ramond field:
  ($g_{\mu\nu}+\beta_{\mu\nu}$).  We construct the  four  dimensional
  covariant  field  theory  and  compute the  resulting  equations  of
  motion.  The equations  encode the  most general form of EMR within
  a well  defined  variational principle,  for specific  lower
  dimensional  embedded   geometric  scenarios. Our formalism also
  reveals the emergence of additional diffusive pseudo currents for a
  completely dynamic field theory of EMR.   The proposed equations of motion now include terms that 
  induce geometrical deformations in the device
  geometry in order to optimize the EMR.  This  bottom-up dual  description between EMR and
generalized geometry/gravity lends itself to a deeper insight into
the EMR effect with the promise of potentially new physical phenomena and properties.
\end{abstract}
\maketitle
\section{Introduction}
Since   the   advent  of   the  Anti-deSitter/Conformal Field Theory
(AdS/CFT)   correspondence  of   string 
theory \cite{Maldacena:1997re}, the  implementation of  gravity duality
theories has seen a plethora of use for applications ranging from black
hole thermodynamics in quantum gravity  to phase transitions and their
critical temperatures in condensed matter. The general idea of mapping
strong-coupling  problems to a weakly  coupled gravity theory,
where perturbative methods and/or analytic approaches are feasible, has
spawned a relatively  new paradigm in condensed
matter physics \cite{Sachdev:2010ch}.

The critical  phenomena related  to phase  changes in  strong coupling
theory map  back to  black hole  thermodynamic properties  of specific
solution  spaces  of  the  dual  gravity  theory.   This  permits  the
application of  known perturbative techniques of  a higher dimensional
gravity  theory  in  situations  where  field-theoretic  or  numerical
methods fail  for specific time-domains and/or  critical temperatures.
Conversely, certain classes of black hole solutions exhibit asymptotic
two  dimensional  chiral  conformal symmetries  which  generate  their
near-horizon structures.   These symmetries allow for  the computation
of certain  quantum gravitational  properties of the  respective black
hole within  known renormalizable  CFT techniques,  thus circumventing
the ultraviolet behavior of four dimensional general relativity.

The discovery  of the extraordinary magnetoresistance  (EMR) effect in
hybrid semiconductor-metal  structures when  cast in the  framework of
action  integral  formulation  provides   another  venue  towards  the
realization of  the gravity-condensed matter interrelation.  Though we
should be careful with use of the word gravity, as our construction is
not   a   traditional   CFT/gravity   correspondence;   however,   the
interrelation described  in this work is more reminiscent of an
analogue gravity system\cite{Golan:2018tdy,UinHe,doi:10.1142/4861,Minic:2008an,Xu:2010eg,Barcelo:2005fc}. We
also note that our analogue model is rooted in a generalized
geometric\cite{Vysoky:2015psz,Jurco:2015xra} construction, which
naturally encodes diffeomorphisms in conjunction with a stringy
Kalb-Ramond two-form symmetry,  and
resulting conserved two-form  currents relating to internal dynamical
magnetic fields. Similar discoveries have been noted in the (differing but seemingly closely related) application of generalized global
one-form symmetry and resulting conserved two-form currents in order to understand dissipative magnetohydrodynamics\cite{Grozdanov:2016tdf}, holographic duals of specific strongly interacting plasmas\cite{Grozdanov:2017kyl}, generalized elasticity theory\cite{Grozdanov:2018ewh} and (more recently) holographic descriptions of the stable quantum matter phases (fracton states) via spin two $U(1)$ gauge field coupled to emergent massless spin two states (gravity)\cite{Pretko:2017fbf,Yan:2019bim}. 

The EMR phenomenon is very sensitively dependent upon the position and width of the voltage and current ports, and more importantly; the device geometry. The EMR can be optimized by inducing geometrical deformations in the device geometry. The main goal of our duality construction is to provide a mathematical framework to address geometric/shape optimization of the device geometry.\cite{ram3d,ram2d2,ram2d1}

As mentioned above, our construction is not strictly tied to low energy
string theory and originates from the semiconductor side of the
duality and draws upon specific semiconductor experimental constructs
and results. 
The  geometric  nature of  the  correspondence  is realized  from  the
previous theoretical  developments. In this setting, the conductivity
tensor is seen  to behave  similar  to  a metric  deformation  of a
generalized geometric (Courant) algebroid \cite{GualtieriGG}, a
formalism which studies the Lie algebra of smooth sections consisting
of direct sums of vectors and co-vectors on smooth
manifolds. Following Ref.~\onlinecite{Vysoky:2015psz} and
Ref.~\onlinecite{Jurco:2015xra} on generalized geometry, we are able
to construct a fully dynamical theory  of all constituent fields of
the EMR  effect. This  bottom-up dual  description, between EMR and
generalized geometry/gravity, lends itself to a deeper insight into
the EMR effect with potentially new physical phenomena and properties
yet to be discovered. 

In Sec.~\ref{Sec:Duality}, we briefly consider the phenomenon of EMR and show the van der Pauw configuration for measuring the EMR. The variational integral representation of the governing equation of motion provides an insight into the linking of EMR with gravity.
This leads to the geometric sector in which the generalization to 4D considerations with gravity redefines the action to include gravity, the electromagnetic and Ramond-Nuevo-Schwarz $U(1)$ symmetry. The new Maxwell equation implies a conserved current that is analogous to other examples of tensor currents discovered in recent works on combining gravity with magnetohydrodynamics, etc. 
Additionally, Sec.~\ref{Sec:Duality} contains all of our main results, as listed below:
\begin{itemize}
\item In Sec~\ref{sec:tta}, the construction of a four dimensional covariant field theory of EMR and computation of all respective Euler-Lagrange equations, for a complete dynamical field content including geometry is done. Thus provides a computational avenue for shape optimization in EMR.
\item In Sec~\ref{sec:cdt}, the discovery of emerging additional diffusive pseudo currents, complementing the fully dynamical field theory of EMR is given.
\item The summary of our constructed duality, is detailed in Table~\ref{tab:ggemr}.
\item In Sec~\ref{sec:21aas}, we break general diffeomorphism symmetry and compute the $(2+1)$ dimensional action and field equations for general axial symmetry. 
\end{itemize}
\noindent The concluding remarks are delivered in Sec.~\ref{Sec:CCRem} and we summarize results and definitions used from the generalized geometry formalism in Appendix~\ref{app:gg}.
\section{Generalized Geometric EMR Duality}
\label{Sec:Duality}
\begin{figure}[t!] 
	\centering
	\begin{subfigure}{0.4\textwidth}
	\includegraphics[width=\textwidth]{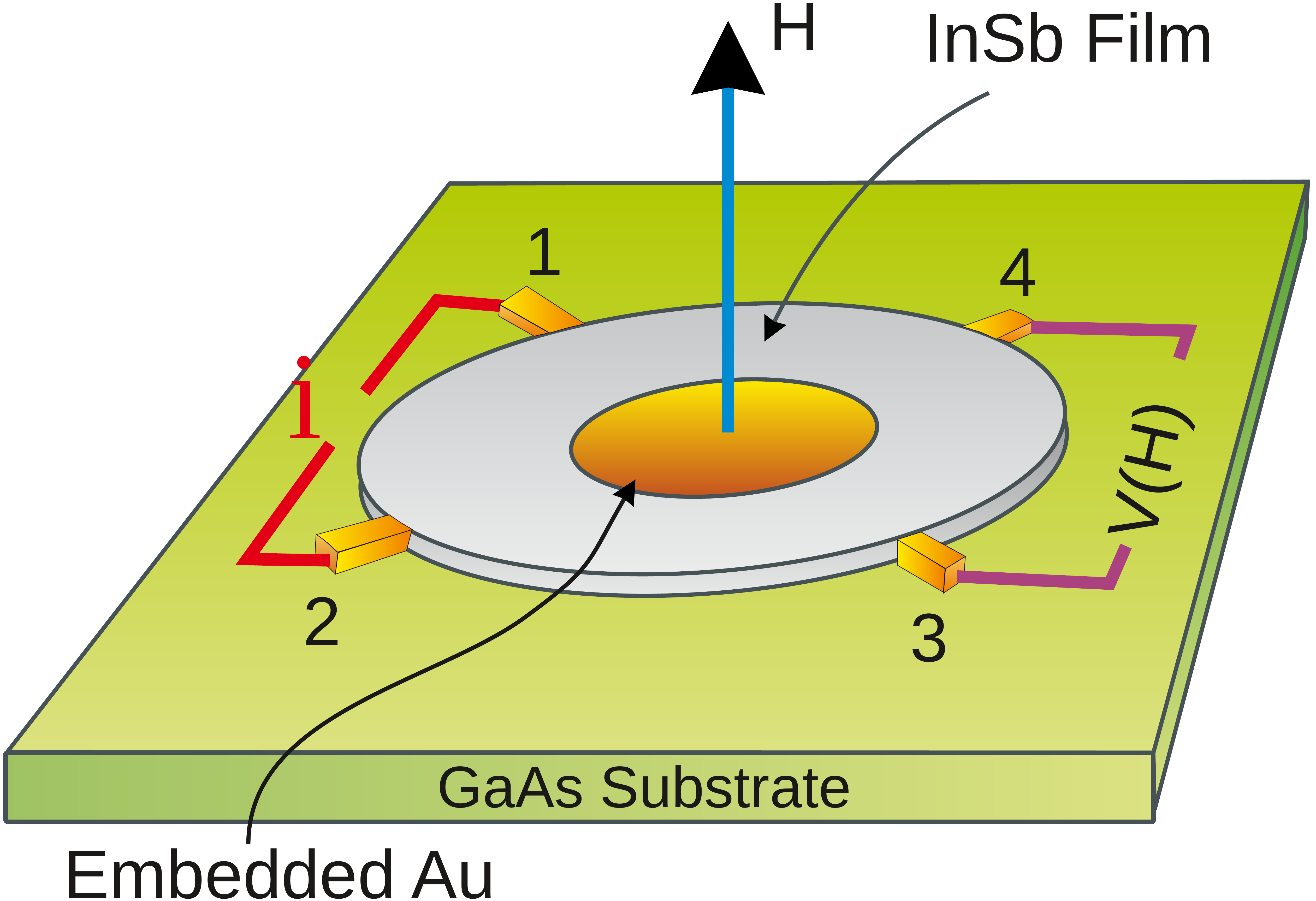}
	\caption{}\label{fig:VanderPauWGeomA}
	\end{subfigure}
	
	\begin{subfigure}{0.4\textwidth}
        \includegraphics[width=\textwidth]{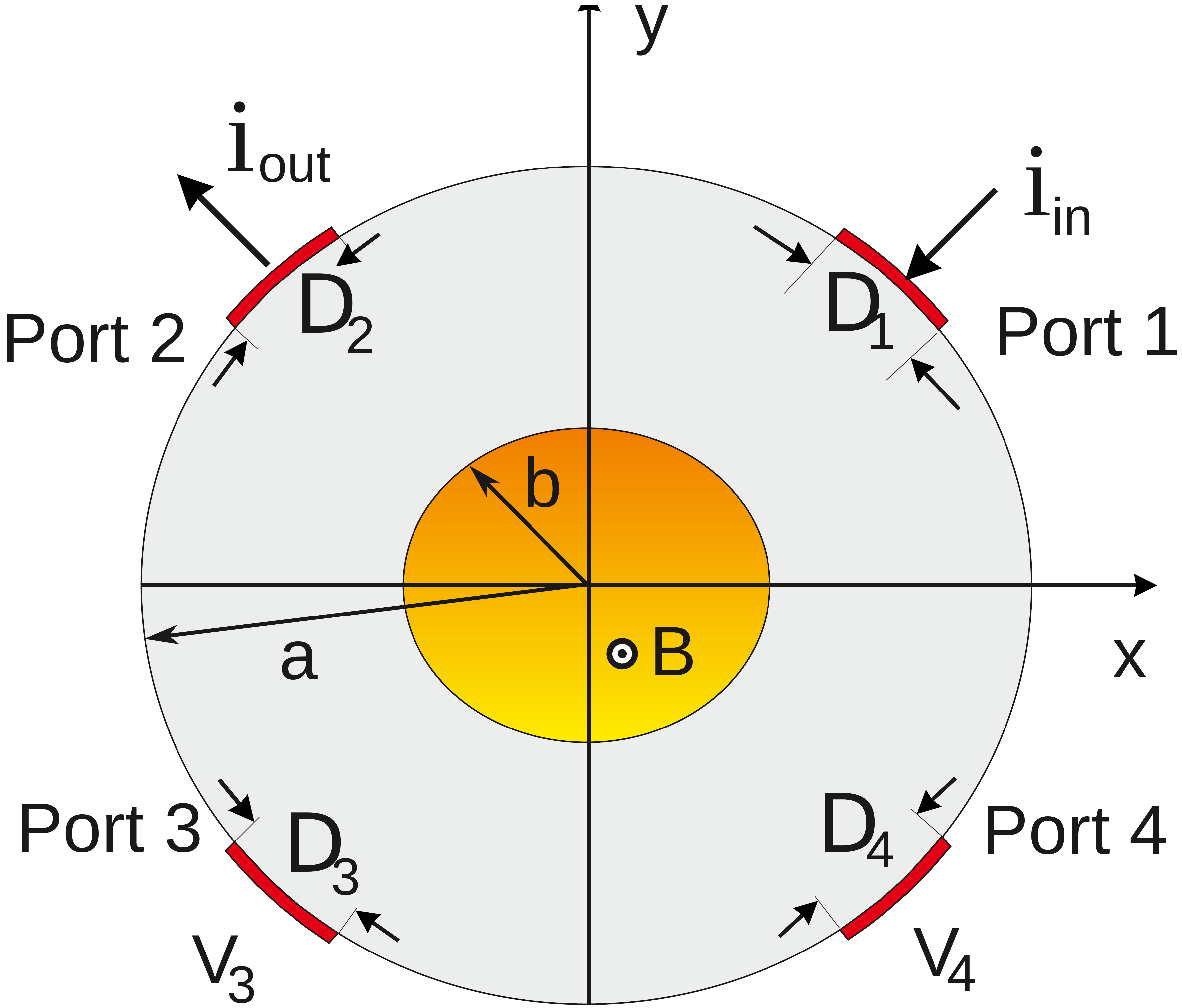}
        \caption{}\label{fig:VanderPauWGeomB}
	\end{subfigure}
 \caption{\label{fig:VanderPauWGeom}A   semiconductor  wafer   with  a
   concentric metallic disk at the center is shown in Fig.~\ref{fig:VanderPauWGeomA}.  In Fig.~\ref{fig:VanderPauWGeomB} The
   4-probe  van der  Pauw arrangement  is displayed  with current  $I$
   entering through port  1 and exiting through port  2.  The contacts
   at ports 3  and 4 are used  to measure the voltage  drop across the
   device. (After Ref.~\onlinecite{ram2d2}).}
      \end{figure}

It has been shown experimentally \cite{Solin1}    that 
semiconductor thin films with metallic  inclusions display EMR at room
temperature,  with  remarkable enhancements  as high  as  100–750000\,\%  at
magnetic fields ranging  from 0.05 to $4\,T$. In the four-probe Hall
measurements, the current through the device is held fixed while the voltage difference
is measured across ports 3 and 4, as in Fig.~\ref{fig:VanderPauWGeom}.
The    magnetoresistance    (MR)     is    defined    as    \mbox{MR$=
  [R(H)-R(0)]/R(0)$}, where  $R(H)$ is the resistance  at finite field
magnetic   field   $H$.    With   a   steady   current   through   the
metal-semiconductor  hybrid  structure,  using  Ohm's  law  we  obtain
$[V(3)-V(4)]_H =  I R(H)$. Thus  the MR  is determined by measuring the
voltage difference across the  device.  The experiments were initially
performed on a  composite van der Pauw disk of  a semiconductor matrix
with an  embedded metallic circular inhomogeneity  that was concentric
with the semiconductor disk.  A  similar enhancement has been reported
\cite{Solin3} for  a rectangular  semiconductor wafer with  a metallic
shunt on one side.  The rectangular geometry with four contacts can be
shown  to be  derivable  from  the circular  geometry  by a  conformal
mapping \cite{Popovic}.  Magnetic  materials and  artificially  layered
metals  exhibit  the  so-called  giant  magnetoresistance  (GMR),  and
manganite perovskites show  colossal magnetoresistance (CMR). However,
patterned nonmagnetic InSb shows  a much larger geometrically enhanced
extraordinary  MR  even  at  room temperature.  This  has  significant
advantages in device design.
\subsection{Condensed Matter Sector}
The MR can be calculated based on a diffusive current-field relation,
\begin{align}\label{eq:olaw}
{\boldsymbol J}=\hat{{\boldsymbol\sigma}}\cdot{\boldsymbol E},
\end{align}
where $\boldsymbol  J$, $\boldsymbol  E$ and $\hat{\boldsymbol\sigma}$ are the usual current density, electric field and conductivity(-tensor). In the presence of an external    (constant)     magnetic    field
${\boldsymbol H}={\boldsymbol  \beta}/\mu$ (where  $\boldsymbol \beta$
 is a unit-less magnetic field and $\mu$ the carrier  mobility)  the (magneto-)conductivity tensor in three dimensional Cartesian coordinates takes the form:
\begin{align}
\begin{split}\label{eq:sig1}
\hat{\boldsymbol\sigma}=&\frac{\sigma_0}{1+|{\boldsymbol \beta}|^2}\times\\
&\hspace{-0.2in}\left(\begin{array}{ccc}\left(1+\beta_x^2\right)     &
    \left(-\beta_z+\beta_y\beta_x\right)                             &
    \left(\beta_y+\beta_z\beta_x\right)
    \\\left(\beta_z+\beta_y\beta_x\right) & \left(1+\beta_y^2\right) &
    \left(-\beta_x+\beta_y\beta_z\right)
    \\\left(-\beta_y+\beta_z\beta_x\right)                           &
    \left(\beta_x+\beta_y\beta_z\right)                              &
    \left(1+\beta_z^2\right)\end{array}\right),
\end{split}
\end{align}
where $\sigma_0$ is the intrinsic conductivity in the presence of zero
external magnetic field. The carrier mobility is given by $\mu=\frac{e\tau}{m^*}$, where $e$ is the electron charge, $\tau$ is the momentum relaxation time and $m^*$ is the effective (electron) mass. Thus, the unit-less magnetic field $\beta$, may equivalently be interpreted as the product; $\beta=\omega_c\tau$, where $\omega_c$ is the effective cyclotron frequency of carriers with mass $m^*$\cite{ram2d1,ram2d2,PhysRevB.49.16256}. In 2D, with ${\boldsymbol H}=\hat{z}H$ and
$\beta_z=\mu H$, we can reduce Eq.~\eqref{eq:sig1} to
\begin{align}
\begin{split}\label{eq:sig1_2D}
\hat{\boldsymbol\sigma}=&\frac{\sigma_0}{1+{\beta_z}^2}
\left(\begin{array}{cc}1 & -\beta_z \\ \beta_z & 1\end{array}\right),
\end{split}
\end{align}
where only the 2D ($x$ and $y$) components are present.

A fast, robust and convergent variational approach for calculating the
EMR for specific geometric cases in 2D and 3D structures has been
developed through the variation of the action integral within the framework of finite
element analysis \cite{ram3d,ram2d2,ram2d1}.
Using the current continuity condition
\begin{align}\label{eq:ccc}
{\boldsymbol\nabla}\cdot\mathbf{J}=0,
\end{align}
we obtain the scalar field equation 
\begin{align}\label{eq:olfeq}
-{\boldsymbol\nabla}\cdot\left(\hat{\boldsymbol\sigma}\cdot{\boldsymbol\nabla}\varphi\right)=0,
\end{align}
where ${\boldsymbol E}=-{\boldsymbol\nabla} \varphi$ and $\varphi$ is the electric
scalar potential.
The action integral for which Eq.~\eqref{eq:olfeq} is the corresponding Euler-Lagrange equation is given by:
\begin{align}
\begin{split}\label{eq:a0action}
{\cal{A}}_0=&\frac12\int d^4x({\boldsymbol\nabla}\varphi)\cdot{\hat{\boldsymbol\sigma}}\cdot({\boldsymbol\nabla}\varphi)\\
=&\frac12\int d^4x\left\{\sum_{i=1}^{3}\sum_{j=1}^{3}\sigma^{ij}\partial_i\varphi\partial_j\varphi\right\}\\
=&\frac12\int d^4x\left(\partial_i\varphi\right)\sigma^{ij}\left(\partial_j\varphi\right),
\end{split}
\end{align}
where we have introduced index notation, and Einstein summation
convention is used throughout this paper for repeated indices omitting the
summation symbol. Latin indices will be reserved for spacial
components and run from $1,2,3$, while greek indices will run from
$0,1,2,3$; where 0 denotes the temporal component. In the case of
a four-probe Hall measurement, to measure MR in a composite structure,
two additional surface terms are added to the action functional to
specify the incoming and outgoing currents.

Looking at Eq.~\eqref{eq:a0action}, we see that by recasting our initial
diffusive equation into a variational form, the conductivity tensor now
plays the role of a matrix encoding geometry, commonly referred to as
the inverse metric of the gravitational field \cite{wald,carroll}. We can now
proceed with a functional variational approach to geometric
optimization by constructing an appropriate action involving the
conductivity tensor given in Eq.~\eqref{eq:sig1}.

The resistivity, denoted as $\tilde{\mathcal{G}}_{ij}$, is given by the inverse of
Eq.~\eqref{eq:sig1}. This yields:
\begin{align}\label{eq:3metric}
\tilde{\mathcal{G}}_{ij}=\frac{1}{\sigma_0}\left[g_{ij}+\beta_{ij}\right],
\end{align}
where $g_{ij}$ is a 3D metric tensor equal to a diagonal matrix, with not necessarily equal diagonal components, depending on choice of coordinates and
\begin{align}\label{eq:KR3field}
  \beta_{ij}=\left(\begin{array}{ccc}0 & \beta_z & -\beta_y \\-\beta_z &
 0 & \beta_x \\ \beta_y & -\beta_x & 0\end{array}\right).
\end{align}

 Further,   we wish to  recast Eq.~\eqref{eq:a0action} into a covariant form (invariant under general
coordinate  transformations). First,  we introduce  the
electromagnetic field curvature tensor, which in Cartesian coordinates
is given by
\begin{align}
F_{\mu\nu}=\left(\begin{array}{cccc}0 & -E_x & -E_y & -E_z \\E_x & 0 &
 -B_z & B_y \\E_y & B_z & 0 & -B_x \\E_z & -B_y & B_x & 0\end{array}\right),
\end{align}
and relates to the gauge-field four vector potential $A_{\mu}=\left(-\varphi,\boldsymbol A\right)$, which is a combination of both
the electric scalar ($\varphi$) and magnetic vector ($\boldsymbol A$) potentials in a four vector
formalism, via
\begin{align}
F_{\mu\nu}=&\partial_\mu A_\nu-\partial_\nu A_\mu.
\end{align}
Clearly,
\begin{align}
\begin{split}\label{eq:mfdef}
  F_{\mu\nu}=&-F_{\nu\mu}\\
  F_{0i}=&\left(\partial_t\boldsymbol
    A+{\boldsymbol\nabla}\varphi\right)_i=-E_i\\
  F_{ij}=&\partial_iA_j-\partial_jA_i=\epsilon\indices{_i_j^k}B_k,
\end{split}
\end{align}
in units where the speed of light is given by $c=1$ and where $\epsilon\indices{_i_j^k}$ is the Levi-Civita symbol, or the totally antisymmetric permutation symbol:
\begin{align}
\epsilon\indices{_i_j^k}=
\begin{cases}
+1&(i,j,k)=(1,2,3)_{even~permutations}\\
-1&(i,j,k)=(1,2,3)_{odd~permutations}\\
0&any~i=j,i=k,j=k
\end{cases}.
\end{align}
Next, we introduce the four dimensional resistivity
$\tilde{\mathcal{G}}_{\mu\nu}=({1}/{\sigma_0})\left[g_{\mu\nu}+\beta_{\mu\nu}\right]$,
where $\tilde{\mathcal{G}}_{ij}$ is as defined in
Eq.~\eqref{eq:3metric}, and the added temporal components are
\begin{align}
\begin{split}\label{eq:sig4def}
\tilde{\mathcal{G}}_{00}=-\frac{1}{\sigma_0},~
\tilde{\mathcal{G}}_{i0}=\sigma_{0i}=0
\end{split}
\end{align}
and the inverse condition reads:
\begin{align}
\begin{split}\label{eq:sigin4def}
\tilde{\mathcal{G}}_{\mu\alpha}\sigma^{\alpha\nu}=&\  \delta\indices{_\mu^\nu}.
\end{split}
\end{align}
Hence, following from Eq.~\eqref{eq:sig1}, \eqref{eq:3metric}, \eqref{eq:sigin4def} we have
\begin{widetext}
\begin{align}\label{eq:Gtss}
\begin{split}
  \tilde{\mathcal{G}}_{\mu\nu}=&\frac{1}{\sigma_0}\left(\begin{array}{cccc}-1
 & 0 & 0 & 0
 \\0 & 1 &
 \beta_z & -\beta_y \\0 & -\beta_z & 1 & \beta_x \\0 & \beta_y &
 -\beta_x & 1\end{array}\right)
\end{split}
\end{align}
and
\begin{align}
\begin{split}
  \sigma^{\mu\nu}=&\frac{\sigma_0}{1+|{\boldsymbol
      \beta}|^2}\left(\begin{array}{cccc}-\left(1+|{\boldsymbol
          \beta}|^2\right) & 0 & 0 & 0 \\0 &\left(1+\beta_x^2\right) &
      \left(-\beta_z+\beta_y\beta_x\right)                           &
      \left(\beta_y+\beta_z\beta_x\right)\\0
      &\left(\beta_z+\beta_y\beta_x\right)  & \left(1+\beta_y^2\right)
      &                        \left(-\beta_x+\beta_y\beta_z\right)\\0
      &\left(-\beta_y+\beta_z\beta_x\right)                          &
      \left(\beta_x+\beta_y\beta_z\right)                            &
      \left(1+\beta_z^2\right)\end{array}\right).
\end{split}
\end{align}
\end{widetext}
With these definitions in hand we may recast Eq.~\eqref{eq:a0action}
into a covariant form via the transformations $d^4x \to
d^4x\sqrt{-g}$, $\mathbf{E}$ and
$\mathbf{B} \to F_{\mu\nu}$ yielding the action
\begin{align}
\begin{split}\label{eq:maxact}
  S_{{\cal{A}}_0}=&-\frac{\lambda}{4}\int d^4x\sqrt{-g}
  \sigma^{\mu\nu}\sigma^{\alpha\beta}F_{\mu\alpha}F_{\nu\beta},
\end{split}
\end{align}
where $\lambda$ is an arbitrary coupling to be determined by requiring
Eq.~\eqref{eq:maxact} to reduce to Eq.~\eqref{eq:a0action} in the
steady state and $g=\det\left(g_{\mu\nu}\right)$. To see this, we may
expand Eq.~\eqref{eq:maxact} in steps ($\mu$ and $\nu$ first),
yielding
\begin{align}
\begin{split}
&S_{{\cal{A}}_0}=-\frac{\lambda}{4}\int d^4x\sqrt{-g}\\
&\left\{\sigma^{00}\sigma^{\alpha\beta}F_{0\alpha}F_{0\beta}+
  2\sigma^{0i}\sigma^{\alpha\beta}F_{0\alpha}F_{i\beta}+
  \sigma^{ij}\sigma^{\alpha\beta}F_{i\alpha}F_{j\beta}\right\}.
\end{split}
\end{align}
The middle term vanishes due to the definitions in Eq.~\eqref{eq:mfdef} and Eq.~\eqref{eq:sig4def} and the above equation reduces to 
\begin{align}
\begin{split}
  &S_{{\cal{A}}_0}=\\
  &-\frac{\lambda}{4}\int
  d^4x\sqrt{-g}\left\{\sigma^{00}\sigma^{\alpha\beta}F_{0\alpha}F_{0\beta}+
\sigma^{ij}\sigma^{\alpha\beta}F_{i\alpha}F_{j\beta}\right\}. 
\end{split}
\end{align}
Next, expanding in $\alpha$ and $\beta$ we obtain
\begin{align}
\begin{split}
&S_{{\cal{A}}_0}=-\frac{\lambda}{4}\int d^4x\sqrt{-g}\left\{\sigma^{00}\sigma^{00}F_{00}F_{00}\right.\\
&\left.+2\sigma^{00}\sigma^{0i}F_{00}F_{0i}+\sigma^{00}\sigma^{ij}F_{0i}F_{0j}+\sigma^{ij}\sigma^{00}F_{i0}F_{j0}\right.\\
&\left.+2\sigma^{ij}\sigma^{0r}F_{i0}F_{jr}+\sigma^{ij}\sigma^{rs}F_{ir}F_{js}\right\},
\end{split}
\end{align}
which, again implementing previous definitions, simplifies to:
\begin{align}
\begin{split}
  S_{{\cal{A}}_0}=&-\frac{\lambda}{4}\int d^4x\sqrt{-g}
  \left\{2\sigma^{00}\sigma^{ij}F_{0i}F_{0j}+
    \sigma^{ij}\sigma^{rs}F_{ir}F_{js}\right\},\\
  =&-\frac{\lambda}{4}\int d^4x\sqrt{-g}\sigma^{00}
  \left\{2\sigma^{ij}E_iE_j-\frac{2\sigma_0g^{ij}B_iB_j}{1+
      |{\boldsymbol \beta}|^2}\right\}.
\end{split}
\end{align}
Now we can solve for the parameter $\lambda$ by enforcing the above equation to
reduce to Eq.~\eqref{eq:a0action} in the steady state ($B_i=0$) and
where $-\partial_i\varphi=E_i$, thus yielding:  
\begin{align}
\lambda=-\frac{1}{\sigma^{00}}. 
\end{align}
\subsection{Geometric Sector}
The overarching goal is to optimize  shape or geometry of the metallic
insulator/semiconductor such that  we obtain desired enhancements to
the EMR.  This can  be  best  accomplished  within an  action
principle,  where  the functional variation of the action yields the
Euler-Lagrange equations whose solutions are  stationary points.  The
task now  is to construct an additional terms to the  action
integral
that  determines the geometric  shape.
We interpret
$\tilde{\mathcal{G}}_{\mu\nu}$  defined  in Eq.(\ref{eq:Gtss}) as   a   metric deformation   of  a
generalized   geometric  structure (algebroid)
\cite{Vysoky:2015psz}. In  other words, $\tilde{\mathcal{G}}_{\mu\nu}$
is the weighted sum of two fields, $g_{\mu\nu}$ which is symmetric and
defines  geometry, and  $\beta_{\mu\nu}$  which  is antisymmetric  and
encodes the components of the  external magnetic field.  The remaining
task now is to  construct an action for $\tilde{\mathcal{G}}_{\mu\nu}$
by looking to string theory as a guide, and implementing techniques of
generalized geometry as reviewed in Appendix~\ref{app:gg}.  

The metric deformation:
\begin{align}
\tilde{\mathcal{G}}_{\mu\nu}=e^{2\psi}\left(g_{\mu\nu}+\beta_{\mu\nu}\right),
\end{align}
includes three fields; $g_{\mu\nu}$ a symmetric Riemannian metric,
$\beta_{\mu\nu}$ akin to the antisymmetric Kalb-Ramond field and the
dilation $\psi$.\cite{Szabo:2002ca} By Riemannian metric, we imply a
metric $g_{\mu\nu}$ that is covariantly constant: 
\begin{align}
\nabla_\alpha g_{\mu\nu}=0,
\end{align}
with respect to the covariant derivative $\nabla_\alpha$, such that:
\begin{align}\label{eq:paratran}
\left(\nabla_\alpha-\partial_\alpha\right)A^\beta=\Gamma^\beta_{\alpha\mu}A^\mu.
\end{align}
In other words, for auto parallel transport of some vector $A^\mu$
across a smooth manifold, the respective change in $A^\mu$ is measured
by the Levi-Civita connection (unique, symmetric and metric compatible\cite{nak}) coefficients $\Gamma^\beta_{\alpha\mu}$,
determined by its Christoffel variant: 
\begin{align}\label{eq:chrissym}
\Gamma\indices{^\alpha_{\mu\nu}} = \frac{1}{2}g^{\alpha\gamma}\left(\partial_\mu g_{\nu\gamma}+\partial_\nu g_{\mu\gamma}-\partial_\gamma g_{\mu\nu}\right).
\end{align}
Thus from Eq.~\eqref{eq:paratran}, we see that the covariant derivative $\nabla_\alpha A^\beta=\partial_\alpha A^\beta+\Gamma^\beta_{\alpha\mu}A^\mu$ is symmetric with respect to general diffeomorphisms.\cite{RRBook} 
The Kalb-Ramond field can simplistically be interpreted as a rank two antisymmetric gauge-field potential, $\beta_{\mu\nu}=-\beta_{\nu\mu}$, with curvature field strength given by its exterior (completely antisymmetric) derivative:
\begin{align}\label{eq:RNschF}
\begin{split}
d\beta=&\mathcal{H},\\
\Rightarrow \mathcal{H}_{\mu\nu\alpha}=&\partial_{\mu}\beta_{\nu\alpha}+
\partial_{\nu}\beta_{\alpha\mu}+\partial_{\alpha}\beta_{\mu\nu}.
\end{split}
\end{align}
This is analogous to the electromagnetic field strength tensor (see Eq.~\eqref{eq:mfdef}) $F_{\mu\nu}=\partial_\mu A_\nu-\partial_\nu A_\mu$, coming from the exterior derivative of the rank one gauge-field potential $A_\mu$. 

One possible route, in constructing a geometric sector action, may be taken by reinterpreting the above three fields as background fields interacting with a closed bosonic string. This scenario is commonly known as a 2 dimensional non-linear sigma model with a world sheet action $S=S_1+S_2+S_3$ comprised of the standard three parts:
\begin{align}\label{eq:stringWS}
\begin{cases}
S_1=& -{\displaystyle{\frac{1}{4\pi\alpha'}}}\int d^2\xi\sqrt{h}h^{ab}\partial_aX^\mu\partial_bX^\nu g_{\mu\nu}(X)\\
\\[0.1pt]
S_2=& -{\displaystyle{\frac{1}{4\pi\alpha'}}}\int d^2\xi\epsilon^{ab}\partial_aX^\mu\partial_bX^\nu \beta_{\mu\nu}(X)\\
\\[0.1pt]
S_3=&{\displaystyle{ \frac{1}{4\pi}}}\int d^2\xi\sqrt{h}R^{(2)}\psi(X)
\end{cases},
\end{align}
where $\alpha'$ is the string coupling,
$\xi^a=\left\{\tau,\sigma\right\}$ are the world sheet coordinates
(parameters), $h_{ab}$ is the world sheet metric, $X^{\mu}$ are the
target spacetime coordinates, $g_{\mu\nu}(X)$ is the target spacetime
metric as a function of $X^\mu$, $\beta_{\mu\nu}(X)$ is the
Kalb-Ramond field as a function of $X^\mu$, $R^{(2)}$ is the induced
Ricci scalar curvature of the string world sheet, and $\psi(X)$ is the
dilation field as a function of $X^\mu$. We should note that $S_3$
above breaks the desired classical conformal invariance of the
respective 2 dimensional non-linear sigma model. However, a standard
renormalization group flow analysis\cite{Callan:1985ia} restores
conformal symmetry at the quantum field theoretic level by
interpreting the resulting one-loop beta functions as field equations,
i.e. 
\begin{align}\label{eq:1loopBF}
\begin{split}
B^g_{\mu\nu}=& R_{\mu\nu}+\frac{1}{4}H\indices{_\mu^{\lambda\rho}}H\indices{_{\nu\lambda\rho}}-2\nabla_\mu\nabla_\nu\psi=0,\\
B^\beta_{\mu\nu}=& \nabla_{\lambda}H\indices{^\lambda_\mu_\nu}-2\nabla_\lambda\psi H\indices{^\lambda_{\mu\nu}}=0,\\
\frac{B^\psi}{\alpha'}=&4\left(\nabla\psi\right)^2-4\nabla_\mu\psi\nabla^\mu\psi+R+\frac{1}{12}H\indices{_{\mu\nu\rho}}H\indices{^{\mu\nu\rho}}=0,
\end{split}
\end{align}
where $R$ is the target spacetime Ricci scalar curvature of
$g_{\mu\nu}$. The above field equations may be derived, up to total
derivatives, via field variations with resect to $g^{\mu\nu}$,
$\beta_{\mu\nu}$ and $\psi$ from the spacetime closed-string effective
action: 
\begin{align}\label{eq:BSupGrav}
S_{\rm eff}=-\frac{1}{2\kappa}\int
  d^{26}x\sqrt{-g}e^{-2\psi}\left\{R-4\left(\nabla\psi\right)^2+\frac{1}{12}H^2\right\}, 
\end{align}
where $H^2=H_{\mu\nu\rho}H^{\mu\nu\rho}$. Note that the above action
requires a spacetime dimension of 26, in order for the conformal
anomaly of $S_3$ to vanish in $B^\psi$\cite{Callan:1985ia}.

Alternatively  for a more symmetric and geometric approach,
and one that is not necessarily rooted/dependent on the 2 dimensional
non-linear sigma model paradigm, we chose to implement the generalized
geometric approach of
Ref.~[\onlinecite{m2019sign}] and [\onlinecite{Lucas:2017idv}] (see
Appendix~\ref{app:gg}) in order to 
construct our action principle. The generalized geometric approach has
also been shown to reproduce the same effective action of
Eq.~\eqref{eq:BSupGrav} by interpreting the generalized Ricci tensor
as a field equation, however for pure symmetry reasons, we are
motivated in interpreting an Einstein-Hilbert action of the
generalized metric as our final geometric sector action. 

For simplicity and continuity with Appendix~\ref{app:gg}, we will
begin with a general metric-deformation given by 
\begin{align}
\mathcal{G}_{\mu\nu}=g_{\mu\nu}+\beta_{\mu\nu}
\end{align}
and after constructing the action for $\mathcal{G}_{\mu\nu}$ it will
be a simple exercise to determine the action for
$\tilde{\mathcal{G}}_{\mu\nu}$ by conformally transforming the action
for $\mathcal{G}_{\mu\nu}$. Looking to diffeomorphism symmetry and
gravity theory/string theory as a guide, we imagine the Euler-Lagrange
equation of motion for a generalized geometric metric to be given by
an Einstein type field equation which is stationary for the respective
generalized Einstein-Hilbert action \cite{LANDAU1975259}: 
\begin{align}\label{eq:gravact}
S_{\mathcal{G}}=\frac{1}{2\kappa^2}\int d^4x\sqrt{-g}\, R,
\end{align}
where $R$ is the Ricci scalar curvature of the general metric
connection $\nabla_{\mathcal{G}}$ and ${1}/{(2\kappa^2)}$  is an
arbitrary coupling. The computation of $R$ in terms of
$\nabla_{\mathcal{G}}$ (here $\nabla$ without a vector symbol denotes
the covariant derivative connection) is detailed in
Appendix~\ref{app:gg} and  reads 
\begin{align}
\begin{split}
R=&R^{LC}-\frac14\mathcal{H}\indices{_\mu_\nu_\alpha}\mathcal{H}\indices{^\mu^\nu^\alpha},\\
=&R^{LC}-\frac14\mathcal{H}^2.
\end{split}
\end{align}
$R^{LC}$  above is  the  Ricci scalar  determined  by the  Levi-Civita
connection coefficients of Eq.~\eqref{eq:chrissym}, such that:
\begin{align}\label{eq:RCTinG}
R^{LC} =g^{\sigma\nu}\left(\partial_\mu\Gamma\indices{^\mu_\nu_\sigma}-\partial_\nu\Gamma\indices{^\mu_\mu_\sigma}+\Gamma\indices{^\mu_\mu_\lambda}\Gamma\indices{^\lambda_\nu_\sigma}-\Gamma\indices{^\mu_\nu_\lambda}\Gamma\indices{^\lambda_\mu_\sigma}\right)
\end{align}  
and $\mathcal{H}$  is determined in Eq.~\eqref{eq:RNschF} in terms of the Kalb-Ramond field $\beta_{\mu\nu}$.

Looking back  at Eq.~\eqref{eq:3metric} we  see that the case  for the
diffusive  current field  relation  is actually  a conformally  scaled
version of $\mathcal{G}$, i.e.,
\begin{align}\label{eq:gct}
\begin{split}
\mathcal{G}_{\mu\nu}&\to \tilde{\mathcal{G}}_{\mu\nu}=e^{2\psi}\mathcal{G}\Rightarrow
\begin{cases}
\tilde{g}_{\mu\nu}=e^{2\psi}g_{\mu\nu}\\
\tilde{\beta}_{\mu\nu}=e^{2\psi}\beta_{\mu\nu}
\end{cases},
\end{split}
\end{align}
where $\psi$ is an arbitrary function, $e^{2\psi}$ is a general
conformal factor and in our case
$e^{2\psi_0}={1}/{\sigma_0}$. Under the above generalized
conformal transformation we have
\begin{align}
\begin{split}
\sqrt{-\tilde{g}}=&e^{4\psi}\sqrt{-g}\\
\tilde{R}^{LC}=&e^{-2\psi}\left[R^{LC}-6\left(\nabla^2\psi+
    \nabla_\mu\psi\nabla^\mu\psi\right)\right],\\
\tilde{\mathcal{H}}^2=&e^{-2\psi}\left[\mathcal{H}^2+
  4\mathcal{H}^{\mu\nu\alpha}\beta_{[\mu\nu}\partial_{\alpha]}\psi+\right.\\
&\left.4\left(\beta_{[\mu\nu}\partial_{\alpha]}\psi\right)
  \left(\beta^{[\mu\nu}\partial^{\alpha]}\psi\right)\right],
\end{split}
\end{align}
and thus Eq.~\eqref{eq:gravact} becomes
\begin{align}\label{eq:gravact1.1}
\begin{split}
  S_{\tilde{\mathcal{G}}}=&\frac{1}{2\kappa^2}\int d^4x\sqrt{-g}
  e^{2\psi}\left\{R^{LC}-18\nabla_{\mu}\psi\nabla^{\mu}\psi\right.\\
  &-\frac{1}{4}\left[\mathcal{H}^2+12\mathcal{H}^{\mu\nu\alpha}
    \beta_{\mu\nu}\nabla_{\alpha}\psi+\right.\\
  &\left.\left.12\left(\beta^2\nabla_\mu\psi\nabla^\mu\psi+
        2\beta\indices{_\mu^\nu}\beta\indices{^\alpha^\mu}
        \nabla_\alpha\psi\nabla_\nu\psi\right)\right]\right\}
\end{split}
\end{align}
up to total derivative terms. Again, from  Eq.~\eqref{eq:3metric} the
above action simplifies drastically for $\psi\to\psi_0$ and
Eq.~\eqref{eq:gravact1.1} becomes
\begin{align}\label{eq:gravact1.2}
\begin{split}
  S_{\tilde{\mathcal{G}}}=&\frac{e^{2\psi_0}}{2\kappa^2}
  \int d^4x\sqrt{-g}\left\{R^{LC}-\frac{1}{4}\mathcal{H}^2\right\}.
\end{split}
\end{align}
Since the dilation field relates to the intrinsic conductivity $\sigma_0$ at zero external magnetic field, there is really no need to consider a dynamical dilation in our action principle, since EMR effects only become present for non-zero external magnetic fields. Thus, the above action $S_{\tilde{\mathcal{G}}}$ is our choice for the geometric sector. 
\subsection{The Total Action}\label{sec:tta}
Collecting results Eq.~\eqref{eq:maxact} and Eq.~\eqref{eq:gravact1.2}
we have the total action
\begin{align}\label{eq:totalaction}
\begin{split}
S_{total}=&S_{\tilde{\mathcal{G}}}+S_{{\cal{A}}_0}\\
=&\frac{e^{2\psi_0}}{2\kappa^2}\int
d^4x\sqrt{-g}\left\{R^{LC}-\frac{1}
  {4}\mathcal{H}^2\right\}\\
&-\frac{\lambda }{4}\int d^4x\sqrt{-g}\sigma^{\mu\nu}
\sigma^{\alpha\beta}F_{\mu\alpha}F_{\nu\beta}.
\end{split}
\end{align}
There are three equations of motion obtained by performing functional
variations with respect to $g^{\mu\nu}$, $A_{\mu}$ and
$\beta_{\mu\nu}$. Implementing the functional variational
relationships
\begin{align}
\begin{split}
  \delta \sigma^{\beta\nu}=&\frac{1}{\sigma_0}\sigma^{\beta\alpha}
  g_{\alpha\lambda}g_{\rho\mu}\sigma^{\mu\nu}\delta g^{\lambda\rho},\\
  \delta \sigma^{\rho\beta}=&-\frac{1}{\sigma_0}\sigma^{\nu\beta}
  \sigma^{\rho\alpha}\delta\beta_{\alpha\nu},
\end{split}
\end{align}
we obtain the following field equations:
\begin{align}\label{eq:eegg}
\begin{split}
&\frac{\delta S_{total}}{\delta g^{\mu\nu}}=0\Rightarrow\\
&\frac{e^{2\psi_0}}{2\kappa^2}\left\{\left(R_{\mu\nu}-\frac12g_{\mu\nu}
    R\right)+\frac18g_{\mu\nu}\mathcal{H}^2-
  \frac34\mathcal{H}_{\mu\alpha\beta}\mathcal{H}\indices{_\nu^{\alpha\beta}}\right\}+\\
 &\frac{\lambda}{4}\left(-\frac12g_{\mu\nu}F^2_\sigma+\frac{2}{\sigma_0}
   \sigma^{\beta\alpha}g_{\alpha\mu}g_{\nu\lambda}\sigma^{\lambda\rho}
   \sigma^{\gamma\epsilon}F_{\beta\gamma}F_{\rho\epsilon}\right)=0,
\end{split}
\end{align}
\begin{align}\label{eq:emrMeq}
\begin{split}
&\frac{\delta S_{total}}{\delta A_{\alpha}}=0\Rightarrow\\
&\nabla_{\mu}\left(\sigma^{\alpha\beta}\sigma^{\mu\nu}F_{\beta\nu}\right)=0,
\end{split}
\end{align}
\begin{align}\label{eq:RNSeq}
\begin{split}
&\frac{\delta S_{total}}{\delta \beta_{\nu\alpha}}=0\Rightarrow\\
&\frac{3e^{2\psi_0}}{2\kappa^2}\nabla_{\mu}\mathcal{H}^{\mu\nu\alpha}+
  \frac{\lambda}{\sigma_{0}}\sigma^{\alpha\rho}\sigma^{\mu\nu}
  \sigma^{\gamma\beta}F_{\mu\gamma}F_{\rho\beta}=0,
\end{split}
\end{align}
where $F^2_{\sigma}=\sigma^{\mu\nu}\sigma^{\alpha\beta}F_{\mu\alpha}F_{\nu\beta}$.
Equation   \eqref{eq:emrMeq}      is   identical  to (or   encodes)
Eq.~\eqref{eq:ccc} and Eq.~\eqref{eq:olfeq} for  $B_i=0$, however now
includes additional dynamics for the  possibility of a non-zero
internal magnetic field. In effect, we have  now enhanced the EMR
geometric optimization problem to  four field equations  above. They
all are  coupled partial differential equations containing geometry
$g_{\mu\nu}$, 
electromagnetism $F_{\mu\nu}$ and external magnetic field
$\beta_{\mu\nu}$. In  the limiting case for  constant $g_{\mu\nu}$ and
$\beta_{\mu\nu}$ all  the curvature  terms involving  $R_{\mu\nu}$ and
$\mathcal{H}_{\mu\nu\alpha}$ are identically  zero and
Eq.~\eqref{eq:eegg} and Eq.~\eqref{eq:RNSeq} reduce to the on-shell
condition of Eq.~\eqref{eq:a0action}. This completes the construction
of our generalized-geometry/EMR duality, which provides a fully
dynamical theory of constituent fields influencing the EMR
phenomena. We should note that the dilation field $\psi$, normally a
dynamical field on the string-theory side, is non-dynamical in our specific
duality. This is due to the material properties of the semiconductor
under the influence of zero external magnetic fields. 

Now, it may be that for a given experiment/design the conductivity tensor and external magnetic field will be ``non-dynamical". However, and as stated earlier, the conductivity tensor (and thus the EMR) depends on the semiconductor shape. Additionally, using time dependent, sinusoidally and radially varying external magnetic fields have become of interest in how they may enhance EMR. The dynamical nature of these fields within our construction allows for the determination, via a variational principle, and selection of the optimum combinations of each field and thus provides parameters for an optimum design for enhancing EMR.
Finally, for this section, we summarize our
duality construction in Table~\ref{tab:ggemr}.  
\begingroup
\squeezetable
\begin{table*}[htbp!]
\caption{\label{tab:ggemr}Dictionary of EMR objects (quantities) and
  their corresponding duals in generalized-geometry (gravity).} 
\begin{ruledtabular}
\begin{tabular}{lrclr}
 \multicolumn{2}{c}{\textbf{Extraordinary-Magnetoresistance}} & \multicolumn{1}{|c|}{}& \multicolumn{2}{c}{\hspace{0.3in}\textbf{Generalized-Geometry (Gravity)}}\\ 
  $A_\mu$ &internal $U(1)$ gauge field &\multicolumn{1}{|c|}{}& $A_\mu$ &externally coupled $U(1)$ gauge field\\
  $\sigma_0$ & intrinsic conductivity at zero external mag. field &\multicolumn{1}{|c|}{}& $\psi_0=-\frac12\ln\sigma_0$ & non-dynamical dilation field (string coupling)\\
  $\mathbf{H}=\frac{1}{\mu}{\boldsymbol\beta}$ & external magnetic field &\multicolumn{1}{|c|}{}& $\beta_{ij}=\epsilon\indices{_i_j^k}\beta_k$ & spacial components of the Kalb-Ramond field\\
  $g_{\mu\nu}$ & semiconductor device geometry &\multicolumn{1}{|c|}{}& $g_{\mu\nu}$ & dynamical metric background (gravity)\\
  $\sigma^{\mu\nu}$ & conductivity tensor &\multicolumn{1}{|c|}{}& $\left(\tilde{\mathcal{G}}_{\mu\nu}=\frac{1}{\sigma_0}\left(g_{\mu\nu}+\beta_{\mu\nu}\right)\right)^{-1}$ & inverse conformal generalized metric\\
\end{tabular}
\end{ruledtabular}
\end{table*}
\endgroup
\subsection{Current Density Tensor}\label{sec:cdt}
The Maxwell  equation, Eq.~\eqref{eq:emrMeq}, implies a
conserved quantity:
\begin{align}
\nabla_{\mu}\left(\sigma^{\alpha\beta}\sigma^{\mu\nu}F_{\beta\nu}\right)=\nabla_{\mu}\left(J^{\alpha\mu}\right)=0,
\end{align}
and thus
\begin{align}\label{eq:emrol}
J^{\alpha\mu}=\sigma^{\alpha\beta}\sigma^{\mu\nu}F_{\beta\nu}.
\end{align}
The above tensor current is antisymmetric $\left(J^{\mu\nu}=-J^{\nu\mu}\right)$ and in the constant $g_{\mu\nu}$ and $\beta_{\mu\nu}$ case:
\begin{align}
J^{0i}=\sigma_0J^i,
\end{align}
where $J^i$ is the electric current in Eq.~\eqref{eq:olaw} and
\begin{align}
J^{ij}=\sigma_0\epsilon\indices{^{ij}_k}J^k_B,
\end{align}
where $J^k_B$ is a pseudo-current and is a new dynamical feature of the EMR/Geometry duality, but is identically zero for zero internal magnetic field $\vec B$. 

The concept of second rank antisymmetric current densities induced by time independent internal magnetic fields was first considered and discussed in \cite{LAZZERETTI1994299}.
A more recent exposition~\cite{CurrentDT} shows that these antisymmetric current densities are fundamental features of materials with internal magnetic fields such as nuclear dipole moments, and in conjunction with magnetizability, nuclear magnetic shielding and spin-spin coupling fully characterize magnetic perturbation responses. A similar analogy can be drawn for our consideration as well, since the pseudo-current here originates from the internal magnetic field. We present a fundamental origin of these currents, within a well defined symmetry principles and the action integral framework. 

An additional interesting analogue including positive and negative magnetoresistive phenomena has been presented in Ref.~[\onlinecite{m2019sign}] and [\onlinecite{Lucas:2017idv}]. In these studies electrons, in clean metals and under certain conditions, are seen to behave collectively as a fluid with magnetohydrodynamic properties, including viscosity, heat transport and entropy. Each of these included properties are accompanied by specific conserved currents which contain pseudo contributions. A comprehensive study of the multi-form symmetries of these currents should yield some interesting theoretical frameworks closley related to the one presented here and the overarching fluid-MHD/gravity correspondence\cite{Rangamani:2009xk,Grozdanov:2016tdf,Grozdanov:2017kyl}.
\subsection{$(2+1)$-Action Ansatz and Equations of Motion}\label{sec:21aas}
In this section we will break diffeomorphism symmetry of Eq.~\eqref{eq:totalaction} in order to construct a model applicable to device designs exhibiting a constant axially symmetric constant external magnetic field. Typically, experiments for measuring EMR are performed on a
semiconductor wafer having a two-dimensional metal semiconductor
hybrid structure. Hence, we present the equations of motion specific
to a 2D configuration space.
We are interested in applying our new EMR formalism to a similar
scenario as analyzed in \cite{ram2d2,ram2d1}. We will need to perform
a dimensional reduction to the total action of
Eq.~\eqref{eq:totalaction} to $2+1$ dimensions, in conjunction with a
cylindrical coordinate choice as in \cite{ram2d2,ram2d1}. We begin
with our symmetric Riemannian metric ansatz, given by: 
\begin{align}
\begin{split}
ds^2=&g_{\mu\nu}dx^\mu dx^\nu\\
=&-f(r)dt^2+f^{-1}(r)dr^2+\ell^2e^{-2\psi(r)}d\varphi^2+dz^2,
\end{split}
\end{align} 
where $\ell$ is an arbitrary length scale. The generalized metric takes the form:
\begin{align}
\begin{split}
  \tilde{\mathcal{G}}_{\mu\nu}=&\frac{1}{\sigma_0}\left(\begin{array}{cccc}-f(r) & 0 & 0 & 0\\
  0 & \frac{1}{f(r)} & r\beta_z & 0 \\
  0 & -r\beta_z & \ell^2e^{-2\psi(r)} & 0\\
  0 & 0 & 0 & 1\end{array}\right)
\end{split},
\end{align}
where we have performed a cylindrical coordinate transformation to Eq.~\eqref{eq:KR3field}, while setting $\beta_z=constant$ and $\beta_x=\beta_y=0$, to coincide with the scenario of \cite{ram2d2,ram2d1}. For these above choices, we have $\mathcal{H}=0$ identically, and the conductivity tensor takes the form:
\begin{widetext}
\begin{align}
\begin{split}
&\sigma^{\mu\nu}=\left(  \tilde{\mathcal{G}}_{\mu\nu}\right)^{-1}=\\
&=\sigma_0\left(\begin{array}{cccc}-{\displaystyle{\frac{1}{f(r)}}} & 0 & 0 & 0\\
  0 & {\displaystyle{\frac{\ell^2f(r)}{\ell^2+e^{2\psi(r)}r^2\beta_z^2f(r)}}} & -{\displaystyle{\frac{e^{2\psi(r)}r\beta_{z}f(r)}{\ell^2+e^{2\psi(r)}r^2\beta_z^2f(r)}}} & 0 \\
  0 & {\displaystyle{\frac{e^{2\psi(r)}r\beta_{z}f(r)}{\ell^2+e^{2\psi(r)}r^2\beta_z^2f(r)} }}& {\displaystyle{\frac{e^{2\psi(r)}}{\ell^2+e^{2\psi(r)}r^2\beta_z^2f(r)}}} & 0\\
  0 & 0 & 0 & 1\end{array}\right).
\end{split}
\end{align}
\end{widetext}
We make the final ansatz for the $U(1)$ gauge field to be given by:
\begin{align}
\begin{split}
A_\mu=\left(-\varphi(r),0,0,0\right).
\end{split}
\end{align} 

Next, we will perform a dimensional reduction of the action
Eq.~\eqref{eq:totalaction} from $(3+1)$ dimensions to $(2+1)$, by
integrating out $z$ from zero to some arbitrary length (height) scale
$L$. This is a convenient choice, since none of the fields in our
theory have explicit $z$ dependence and thus all curvature invariants
in Eq.~\eqref{eq:totalaction} are identical in terms of our ans\"atze
$f(r)$, $\psi(r)$ and $\varphi(r)$. Thus the dimensionally reduced
action is identical in form to Eq.~\eqref{eq:totalaction} modulo an
overall factor of $L$: 
\begin{align}\label{eq:totalaction3D}
\begin{split}
S^{(2+1)}_{total}=&S^{(2+1)}_{\tilde{\mathcal{G}}}+S^{(2+1)}_{{\cal{A}}_0}\\
=&\frac{Le^{2\psi_0}}{2\kappa^2}\int
d^3x\sqrt{-g^{(3)}}R_{(3)}^{LC}\\
&-\frac{L\lambda }{4}\int d^3x\sqrt{-g^{(3)}}\sigma_{(3)}^{\mu\nu}
\sigma_{(3)}^{\alpha\beta}F^{(3)}_{\mu\alpha}F^{(3)}_{\nu\beta}\\
=&\frac{Le^{2\psi_0}}{2\kappa^2}\int d^3x e^{-\psi(r)}\ell\left\{2f'(r)\psi'(r)-2f(r)\psi'(r)^2\right.\\
&\left.-f''(r)+2f(r)\psi''(r)\right\}\\
&-\frac{L\lambda }{4}\int d^3x e^{-\psi(r)}\ell \frac{2\ell^2\sigma_0^2\varphi'(r)^2}{\beta_z^2e^{2\psi(r)}r^2f(r)+\ell^2}
\end{split}
\end{align}
and in terms of the $(2+1)=(3)$ dimensional invariants generated by the effective fields:
\begin{align}
\begin{split}
ds^2=&g^{(3)}_{\mu\nu}dx^\mu dx^\nu\\
=&-f(r)dt^2+f^{-1}(r)dr^2+\ell^2e^{-2\psi(r)}d\varphi^2,\\
A^{(3)}_\mu=&\left(-\varphi(r),0,0\right)
\end{split}
\end{align} 
and
\begin{widetext}
\begin{align}
\begin{split}
&\sigma_{(3)}^{\mu\nu}\\
&=\sigma_0\left(\begin{array}{cccc}-{\displaystyle{\frac{1}{f(r)}}} & 0 & 0 \\
  0 & {\displaystyle{\frac{\ell^2f(r)}{\ell^2+e^{2\psi(r)}r^2\beta_z^2f(r)}}} & -{\displaystyle{\frac{e^{2\psi(r)}r\beta_{z}f(r)}{\ell^2+e^{2\psi(r)}r^2\beta_z^2f(r)} }}\\
  0 & {\displaystyle{\frac{e^{2\psi(r)}r\beta_{z}f(r)}{\ell^2+e^{2\psi(r)}r^2\beta_z^2f(r)}}} & {\displaystyle{\frac{e^{2\psi(r)}}{\ell^2+e^{2\psi(r)}r^2\beta_z^2f(r)}}}\end{array}\right).
\end{split}
\end{align}
\end{widetext}
The dimensionally reduced action has two general equations of motion;
the Einstein field equations, which come from variation with respect
to $g^{\mu\nu}_{(3)}$ and the Maxwell equations, which come from
variations with respect to $A^{(3)}_\mu$: 
\begin{widetext}
	\begin{align}
	\begin{split}
	\frac{\delta S^{(2+1)}_{total}}{\delta g^{\mu\nu}_{(3)}}=0\Rightarrow&\begin{cases}
	{\displaystyle{\frac{e^{2 \text{$\psi $0}} \left(f'(r) \psi '(r)+2 f(r)\left(\psi''(r)-\psi'(r)^2\right)\right)}{\kappa^2}}}+{\displaystyle{\frac{\lambda\sigma_0^2 \ell^2 \varphi'(r)^2}{\beta_z^2e^{2 \psi(r)}r^2 f(r)+\ell ^2}}}&=0\\
	\\[0.1pt]
	{\displaystyle{\frac{\lambda\sigma_0^2\ell^2\left(\beta_z^2e^{2\psi(r)}r^2f(r)-\ell^2\right)\varphi'(r)^2}{\left(\beta_z^2e^{2\psi(r)}r^2f(r)+\ell^2\right)^2}}}-{\displaystyle{\frac{e^{2\psi_0}f'(r)\psi'(r)}{\kappa^2}}}&=0\\
	\\[0.1pt]
	{\displaystyle{\frac{\lambda\sigma_0^2\ell^2\left(3\beta_z^2e^{2\psi(r)}r^2f(r)+\ell^2\right)\varphi'(r)^2}{\left(\beta_z^2e^{2\psi(r)}r^2f(r)+\ell^2\right)^2}}}+{\displaystyle{\frac{e^{2\psi_0}f''(r)}{\kappa^2}}}&=0
	\end{cases}
	\end{split}
	\end{align}
	and
	\begin{align}
	\begin{split}
	\frac{\delta S^{(2+1)}_{total}}{\delta A_{\mu}^{(3)}}=0\Rightarrow&
	\beta_z^2e^{2\psi(r)}r^2f'(r)\varphi'(r)+\ell^2\left(\varphi'(r)\psi'(r)-\varphi''(r)+\beta_z^2e^{2\psi(r)}rf(r)\left(\varphi'(r)\left(2+3r\psi'(r)\right)-r\varphi''(r)\right)\right)=0.
	\end{split}
	\end{align}
\end{widetext}
The above equations are completely new results in the effort to address the effects of dynamical geometrical deformations in optimizing the EMR, and are not obtainable within the diffusive current-field relation, Eq.~\eqref{eq:olaw}, alone. Despite specifying a constant external magnetic field, the metric is still dynamical for the initially specified symmetries.
A general solution to the above dimensionally reduced action and
resulting equations of motion is currently in progress using numerical
techniques within the finite element analysis approach \cite{ram2d1},
with results to follow in a forthcoming manuscript \cite{GGEMR2}.  

As a  proof-of-concept we consider a pure
simplistic geometry mimicking the device geometry depicted in
Fig.~\ref{fig:VanderPauWGeom}, for which
\begin{align}
\begin{split}
f(r)=1,\\
\psi(r)=\frac12\ln{\frac{\ell^2}{r^2}},
\end{split}
\end{align}
\begin{figure}[t!] 
\begin{center} 
	\begin{subfigure}{.4\textwidth}
	\includegraphics[width=\textwidth]{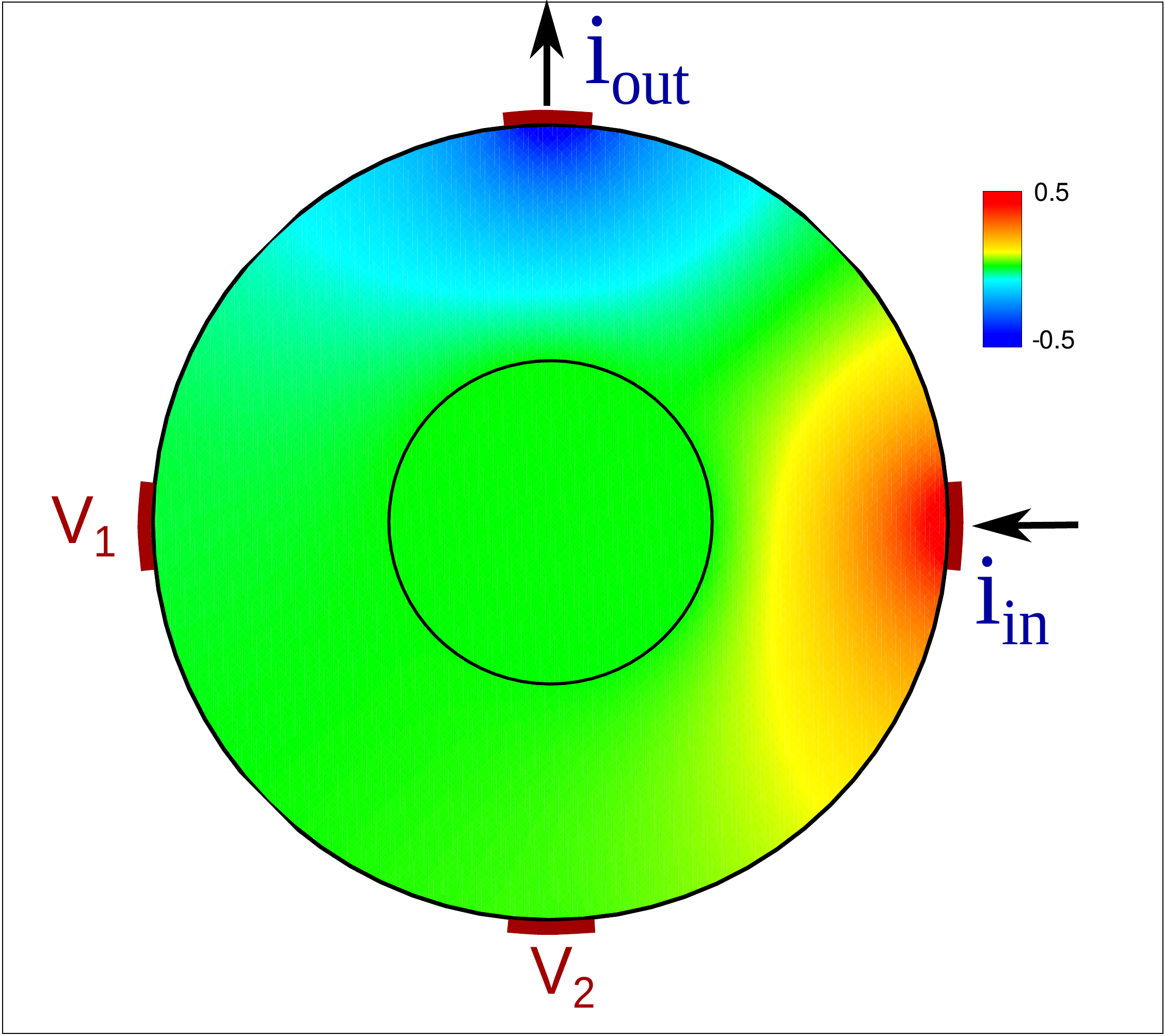}
	 \caption{}\label{fig:FEMB0}
	\end{subfigure}
	\begin{subfigure}{0.4\textwidth}
        \includegraphics[width=\textwidth]{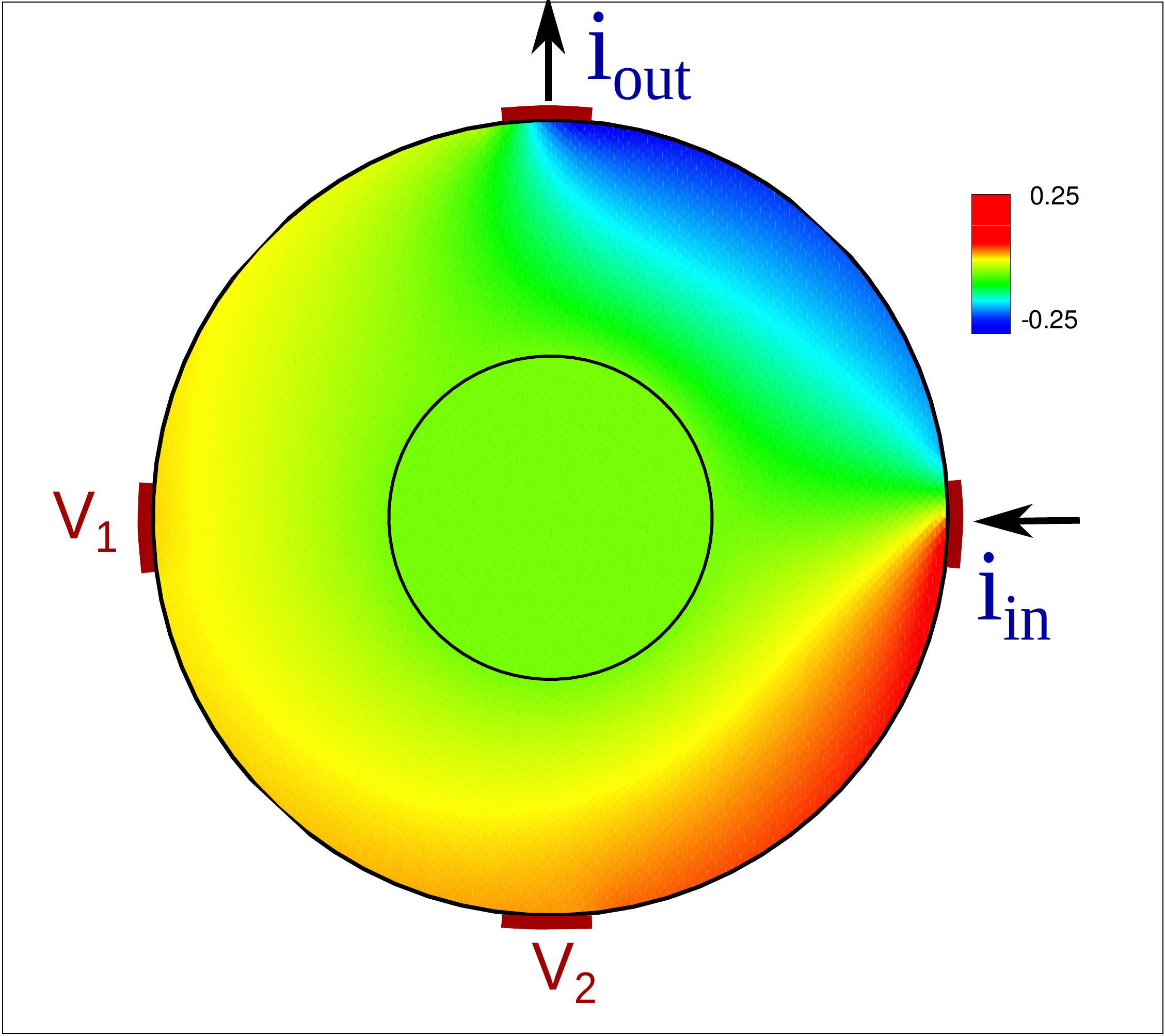}
        \caption{}\label{fig:FEMB5}
	\end{subfigure}
\end{center}
 \caption{\label{fig:FEMPot}We plot the potential function solved for
   the device geometry in Fig.~\ref{fig:VanderPauWGeom}, for two cases
   in which the applied magnetic field (\ref{fig:FEMB0}) \mbox{$H=0\,$T}, and (\ref{fig:FEMB5})
   \mbox{$H=1\,$T}. The color coding from blue to red represents
   the variation of the potential from negative to positive
   values}
\end{figure} 
\begin{figure}[t!] 
\begin{center} 
	\begin{subfigure}{0.4\textwidth}
	\includegraphics[width=\textwidth]{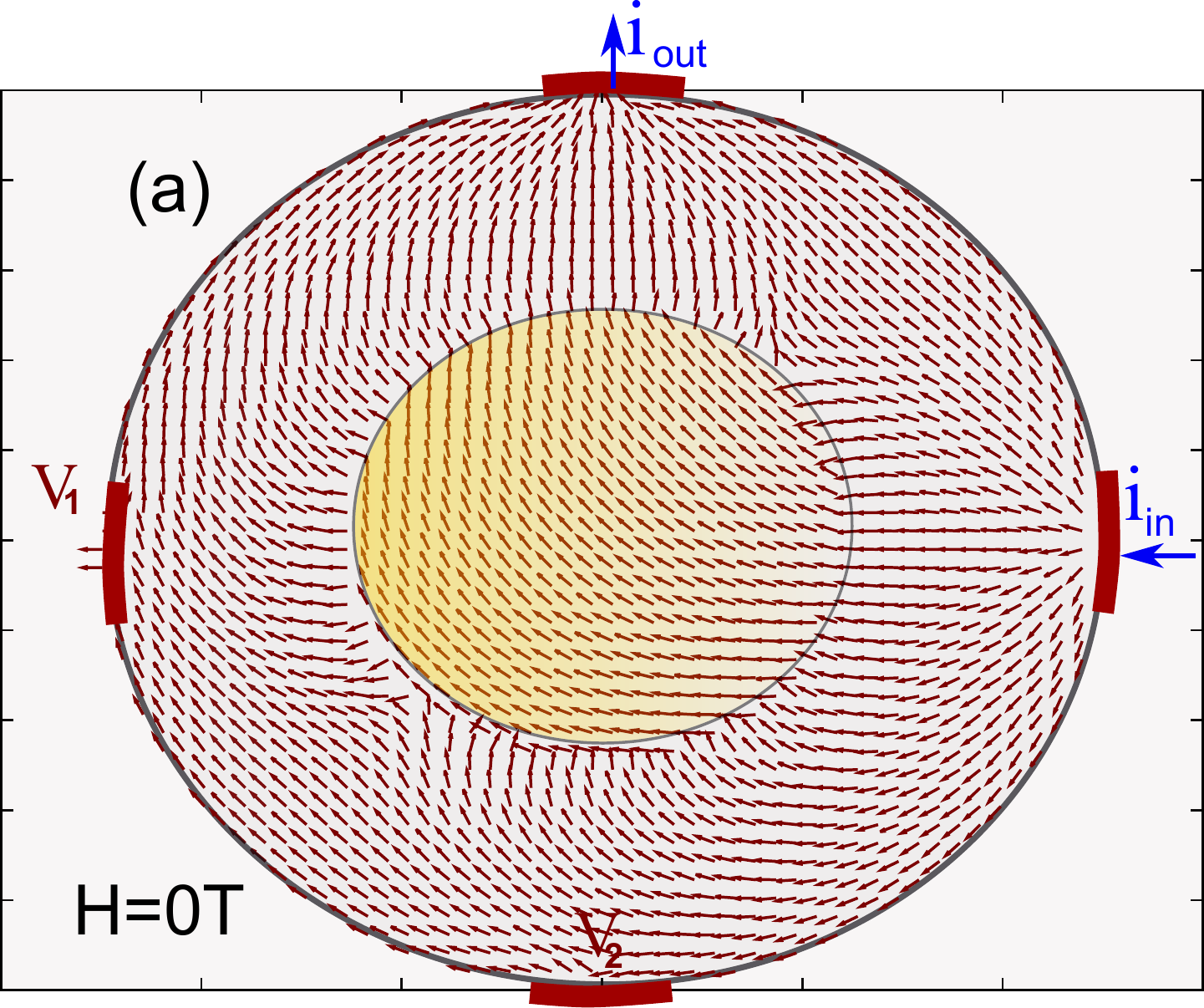}
	 \caption{}\label{fig:FEMcurrentB0}
	\end{subfigure}
	\begin{subfigure}{0.4\textwidth}
        \includegraphics[width=\textwidth]{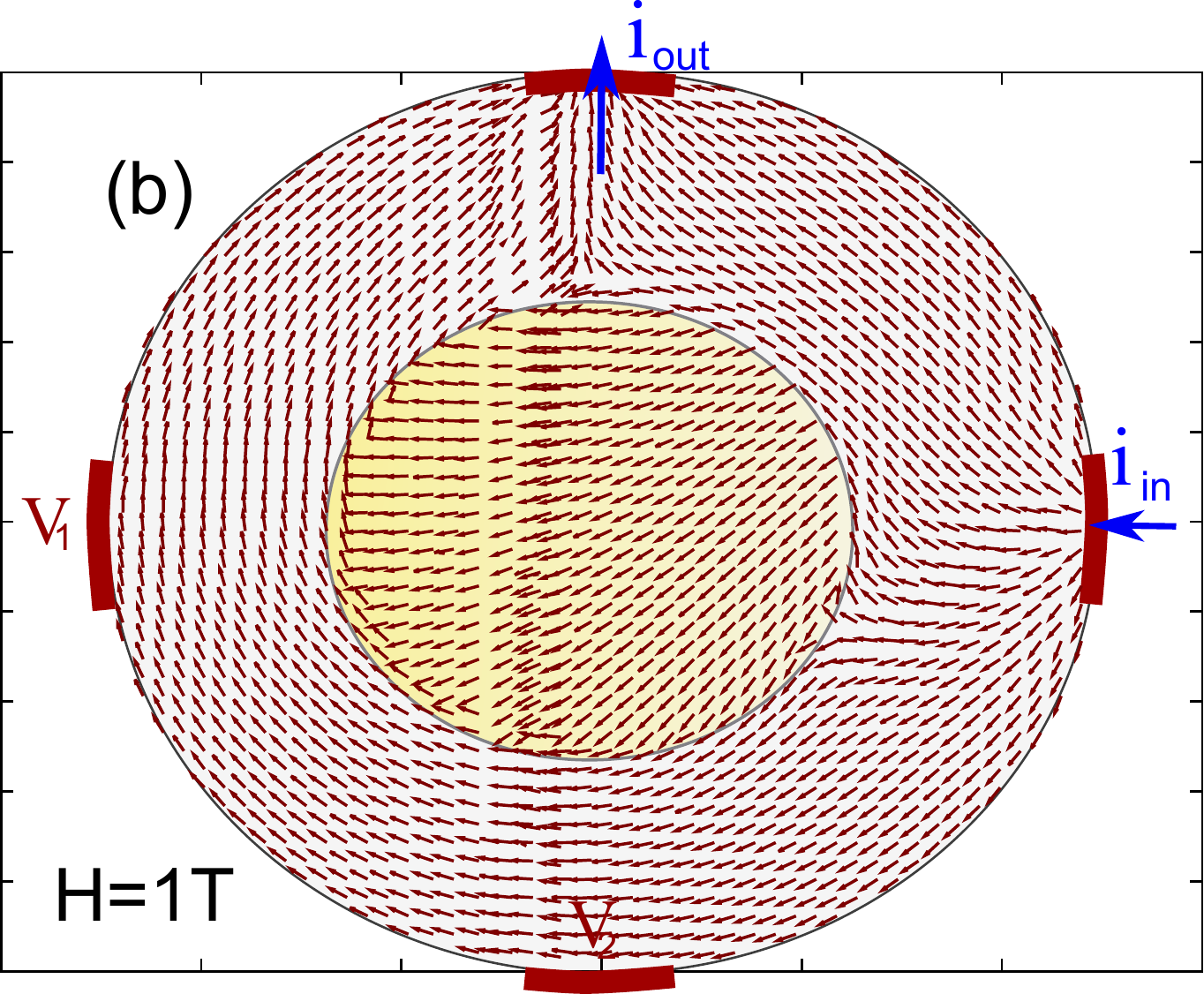}
        \caption{}\label{fig:FEMcurrentB1}
	\end{subfigure}
 \caption{\label{fig:FEMCurrent}We plot the current flow (arrows) for the 
   the device geometry in Fig.~\ref{fig:VanderPauWGeom}, for two cases
   in which the applied magnetic field (\ref{fig:FEMcurrentB0}) \mbox{$H=0\,$T}, and (\ref{fig:FEMcurrentB1})
   \mbox{$H=1\,$T}. Note that the current avoids the metallic region
   at high fields and stays in the semiconductor region, where the Hall angle approaches $\pi/2$, leading to
 extraordinarily high magnetoresistance. The rescaling of the arrows
 everywhere manifests itself as the metallic region carrying a current
even though it is reduced substantially there}
\end{center}
\end{figure} 
and solve the dimensionally reduced action integral with the above
assumptions within the framework of finite element
analysis~\cite{ram2d1}. The device geometry is discretized into a
refined finite element mesh. Within each element, we express the
potential 
function as a linear combination of Hermite interpolation polynomials
multiplied by as-yet undetermined coefficients. Using the principle of
stationary action, variation of the action integral with respect to
these undetermined coefficients results in a set of linear equations,
which are then solved using sparse matrix solvers.  

In Fig.~\ref{fig:FEMB0} and \ref{fig:FEMB5}, we plot the potential
function obtained for the two cases with a constant applied magnetic field
\mbox{$H=0\,$T}, and \mbox{$H=1\,$T}, respectively. We note that for
a non-zero external magnetic field, the potential gradient (current
density) clearly shows that the current is expelled\cite{ram3d} from
the metallic region into the semiconductor region, as shown in
Figs.~\ref{fig:FEMcurrentB0}  and \ref{fig:FEMcurrentB1} and in accordance with Ref.~\onlinecite{ram2d1}.  This is because, while
$\mathbf{E}$ is always perpendicular to the metal surface, at
$\mathbf{H}\neq0$, $\mathbf{J}$ deflects away from $\mathbf{E}$ by the
amount of the Hall angle.  Hence, the magnetoresistance increases
significantly with the applied external magnetic field. With three
orders of magnitude difference between the conductivities of
semiconductor and the metallic inclusin.  The EMR will be very
sensitive to the position and width of the voltage and current ports,
and more importantly to the device geometry. 
\section{Concluding remarks and outlook}\label{Sec:CCRem}
Here, we  have recognized the existence  of a metric deformation  of a
generalized  geometric structure  for the  magnetoconductivity tensor,
that  arises in  the diffusive  model for  the current-electric  field
relation.  We have derived a new action integral and the corresponding
field equations  which encode contributions from  the condensed matter
and the geometrical sector. This is a bottom-up dual description which
provides  a  fundamental  alternative to  the  inter-relation  between
condensed  matter  and CFT/gravity.   Though  we  have motivated  this
approach through 
the EMR effect, our
approach  is  very general  since  we  begin  with the  basic  current
continuity  condition. This  should be  of interest  for a  variety of
geometrical     and     material    optimization     problems     with
semiconductor-metal hybrid structures  in inhomogenous magnetic fields
for wider device applications  such as magento-sensors, readheads, and
the like.

Finally, some immediate questions/comments for future work have arisen:
\begin{itemize}
\item The resulting field equations of Section~\ref{sec:21aas} are currently being analyzed via the finite element analysis paradigm for a specific choice of geometric optimizations, with final results to be presented in a forthcoming publication\cite{GGEMR2}.
\item In our construction, we did not address gauge invariance of the
  Kalb-Ramond field in Eq.~\eqref{eq:totalaction}
  ($\beta_{\mu\nu}$). Requiring this symmetry should induce an
  additional conserved two-form current ($J_\beta^{\mu\nu}$) as a
  response from the condensed matter sector.  
\item The study of the above gauge symmetry might give a relationship to global one-form (or even higher-form) symmetry and their resulting conserved currents, a potentially very interesting relationship that needs to be explored, and thus relating to Ref.~[\onlinecite{Grozdanov:2016tdf,Grozdanov:2017kyl,Grozdanov:2018ewh}].
\item Now that we have a full field theory of EMR, the relatively
  unexplored realm of quantum enhancements of EMR can be explored via
  path integral tree level perturbation theory from the perspective of
  the gravity side. 
\end{itemize}
\begin{acknowledgments}
LR and SR would like to thank Vincent Rodgers, Catherine Whiting and Djordje Minic for
support, encouragement and very enlightening discussions. SB and LRR
thank WPI for the computational resources used in all the
simulations. 
\end{acknowledgments}
\appendix
\section{Generalized Geometry Redux}\label{app:gg}
This section is not intended to provide a full pedagogical introduction to the generalized geometry formalism, but is included for completeness and to demonstrate where the geometric sector action functional Eq.~\eqref{eq:gravact1.2} originates from. For a comprehensive educational introduction we refer to the dissertation works of Gualtieri \cite{GualtieriGG} and Vysoky.\cite{Vysoky:2015psz}

As  mentioned before, the inverse of $\hat{\boldsymbol\sigma}$ in
Eq.~\eqref{eq:3metric} may be interpreted as the metric deformation of
a generalized geometric structure (algebroid).\cite{Vysoky:2015psz} In
this setting and following the notation of
Ref.~[\onlinecite{Vysoky:2015psz}] and
Ref.~[\onlinecite{Jurco:2015xra}] we consider a general bundle
$E=T(\mathcal{M})\oplus T^*(\mathcal{M})$ of the manifold
$\mathcal{M}$, such that
\begin{itemize}
\item $T(\mathcal{M})$ is the tangent bundle, and
\item $T^*(\mathcal{M})$ is the co-tangent bundle.
\end{itemize}
Thus, a smooth section $e$ $\in$ $\Gamma(E)$ is the direct sum
\begin{align}
e=X+\xi,
\end{align}
where $X=X^\mu\partial_\mu$ is a vector and $\xi=\xi_\mu dx^\mu$ is a
1-form. A natural pairing invariant under $O(d,d)$ rotations, where
$d$ is the dimension of $\mathcal{M}$, is given by
\begin{align}
\begin{split}
\left\langle e_1,e_2\right\rangle=&\left\langle X+\xi,Y+\eta\right\rangle\\
&=i_Y\xi+i_X\eta\\
&=Y^\mu\xi_\mu+X^\mu\eta_\mu,
\end{split}
\end{align}
where $i$ denotes the interior product. A bracket structure (similar
to a Lie bracket, but not identical) is given by the Dorfman bracket
\begin{align}
\begin{split}
\left[e_1,e_2\right]_D=[X,Y]_{Lie}+\mathcal{L}_X\eta-i_Yd\xi,
\end{split}
\end{align}
where $[,]_{Lie}$ is the standard Lie bracket between vectors, $\mathcal{L}$ is the Lie derivative and $d$ is the exterior derivative, i.e.,
\begin{align}
\begin{split}
\left([X,Y]_{Lie}\right)^\nu=&X^\mu\partial_\mu Y^\nu-Y^\mu\partial_\mu X^\nu\\
\left(\mathcal{L}_X\eta\right)_\nu=&X^\mu\partial_\mu\eta_\nu+\eta_\mu\partial_\nu X^\mu\\
\left(i_Yd\xi\right)_\nu=&Y^\mu\partial_{[\mu}\xi_{\nu]}.
\end{split}
\end{align}
Next, we define the anchor map $a:E\to T(\mathcal{M})$, as the projection
\begin{align}
a(e)=a(X+\xi)=X.
\end{align}
Now, for calculational purposes, the collection $\left\{E;\langle,\rangle;[,]_D;a\right\}$ forms a Courant algebroid such that the following is specifically satisfied
\begin{itemize}
\item Leibniz Rule\\For all $f$ $\in$ $C^\infty(\mathcal{M})$ the Dorfman bracket satisfies
\begin{align}
\left[e_1,fe_2\right]_D=f\left[e_1,e_2\right]_D+\left(a\left(e_1\right)f\right)e_2;
\end{align}
\item Jacobi Identity
\begin{align}
\left[e_1,\left[e_2,e_3\right]_D\right]_D+\left[e_2,\left[e_3,e_1\right]_D\right]_D+\left[e_3,\left[e_1,e_2\right]_D\right]_D=0;
\end{align}
\item Homomorphism and Leibniz of $a$
\begin{align}
a\left(\left[e_1,e_2\right]_D\right)&=\left[a\left(e_1\right),a\left(e_2\right)\right]_D\\
a\left(e_1\right)\left\langle e_2,e_3\right\rangle&=\left\langle\left[e_1,e_2\right]_D,e_3\right\rangle+\left\langle e_2,\left[e_1,e_3\right]_D\right\rangle\\
a^\dagger d\left\langle e_1,e_2\right\rangle&=\left[e_1,e_2\right]_D+\left[e_2,e_1\right]_D,
\end{align}
where $a^\dagger:T^*(\mathcal{M})\to E^*\sim E$.
\end{itemize}
The $O(d,d)$ symmetries are given by
\begin{align}\label{eq:TGGSym}
T=\left(\begin{array}{cc}N & \beta^* \\\beta & -N^*\end{array}\right),
\end{align}
where
\begin{align}
&N:X^\mu\to N\indices{^\mu_\nu}X^\nu&{\rm Diffeomorphism}\nonumber\\
&\beta:X^\mu\to\beta_{\mu\nu}X^\nu&{\rm Kalb-Ramond~field}\nonumber\\
&\beta^*:\xi_\mu\to \beta^{\mu\nu}\xi_\nu&{\rm Bivector}\nonumber\\
&N^*:\xi_\mu\to N\indices{_\mu^\nu}\xi_\nu&{\rm Diffeomorphism}.
\end{align}
Given the above, a metric deformation may be introduced
\begin{align}
\left\langle e_1,e_2\right\rangle\to&\left\langle e_1,e_2\right\rangle^{\mathcal{G}}\\
&=\left\langle e^{\mathcal{G}}\left(e_1\right),e^{\mathcal{G}}\left(e_2\right)\right\rangle,
\end{align}
and
\begin{align}
\left[ e_1,e_2\right]_D\to&\left[ e_1,e_2\right]_D^{\mathcal{G}}\\
&=e^{-\mathcal{G}}\left[e^{\mathcal{G}}\left(e_1\right),e^{\mathcal{G}}\left(e_2\right)\right]_D,
\end{align}
where $\mathcal{G}_{\mu\nu}=g_{\mu\nu}+\beta_{\mu\nu}$ which maps from
$T(\mathcal{M})\to T^*(\mathcal{M})$ and $e^{\mathcal{G}}:E\to
E$. Specifically we have
\begin{align}
e^{\mathcal{G}}\left(e\right)=e+\mathcal{G}\left(a(e),-\right),
\end{align}
i.e., for $e=X+\xi$ the above reads
\begin{align}
e^{\mathcal{G}}\left(X+\xi\right)=X+\xi+\left(g_{\mu\nu}+\beta_{\mu\nu}\right)X^\nu dx^\mu.
\end{align}
The above definitions imply
\begin{align}
\left\langle e_1,e_2\right\rangle^{\mathcal{G}}=\left\langle e_1,e_2\right\rangle+2g(X,Y),
\end{align}
and
\begin{align}
\left[ e_1,e_2\right]_D^{\mathcal{G}}=\left[ e_1,e_2\right]_D+2g(\nabla X,Y),
\end{align}
where $\nabla$ is the generalized connection of the non-symmetric metric $\mathcal{G}$ and is defined in terms of the general Koszul formula
\begin{align}
\begin{split}
2g(\nabla_Z X,Y)=&X\mathcal{G}(Y,Z)-Y\mathcal{G}(X,Z)+Z\mathcal{G}(X,Y)\\
&-\mathcal{G}\left(Y,\left[X,Z\right]_{Lie}\right)-\mathcal{G}\left(\left[X,Y\right]_{Lie},Z\right)\\
&+\mathcal{G}\left(X,\left[Y,Z\right]_{Lie}\right).
\end{split}
\end{align}
Working out the computational details of the above we see a splitting
in the generalized connection such that
\begin{align}
g\left(\nabla_X Y,Z\right)=g\left(\nabla_X^{LC} Y,Z\right)+\frac12\mathcal{H}(X,Y,Z),
\end{align}
i.e.,  the contortion of $\nabla$ is given by the Ramond-Neveu-Schwarz
three form $\mathcal{H}=d\beta$ and $\nabla^{LC}$ is the Levi-Civita connection
obtained from just $g$. As a result, we obtain the generalized Ricci
tensor
\begin{align}
R_{\mu\nu}=R_{\mu\nu}^{LC}-\frac12\nabla^{LC}_{\alpha}\mathcal{H}\indices{_\mu_\nu^\alpha}-\frac14\mathcal{H}\indices{_\nu_\beta^\alpha}\mathcal{H}\indices{_\alpha_\mu^\beta},
\end{align}
and the Ricci scalar
\begin{align}
R=R^{LC}-\frac14\mathcal{H}\indices{_\mu_\nu_\alpha}\mathcal{H}\indices{^\mu^\nu^\alpha}.
\end{align}


\bibliography{cftgr}

\end{document}